\documentclass[twocolumn,showpacs,aps,epsfig,nofootinbib]{revtex4}

\usepackage{graphicx}
\usepackage{epstopdf}
\usepackage{latexsym}
\usepackage{amssymb}
\usepackage{color}

\usepackage[center]{subfigure}

\begin{document}

 \newcommand{\bq}{\begin{equation}}
 \newcommand{\eq}{\end{equation}}
 \newcommand{\bqn}{\begin{eqnarray}}
 \newcommand{\eqn}{\end{eqnarray}}
 \newcommand{\nb}{\nonumber}
 \newcommand{\lb}{\label}
\newcommand{\PRL}{Phys. Rev. Lett.}
\newcommand{\PL}{Phys. Lett.}
\newcommand{\PR}{Phys. Rev.}
\newcommand{\CQG}{Class. Quantum Grav.}

\title{Holographic Superconductors in Ho\v{r}ava-Lifshitz Gravity}

\author{Kai Lin}
\email{lk314159@hotmail.com}

\author{E. Abdalla}
\email{eabdalla@usp.br}

\affiliation{Instituto de F\'isica, Universidade de S\~ao Paulo, CP
66318, 05315-970, S\~ao Paulo, Brazil}

\author{Anzhong Wang}
\email{Anzhong_Wang@baylor.edu}

\affiliation{GCAP-CASPER, Physics Department, Baylor University,
Waco, TX 76798-7316, U.S.A.}

\date{\today}

\begin{abstract}
We consider holographic superconductors related to the Schwarzschild black hole
in the low energy limit of Ho\v{r}ava-Lifshitz spacetime. The
non-relativistic electromagnetic and scalar fields are 
introduced to construct a holographic superconductor model in
Ho\v{r}ava-Lifshitz gravity and the results show that the
$\alpha_2$ term plays an important role,  modifying the conductivity
curve line by means of an attenuation the conductivity.
\end{abstract}


\maketitle

\section{Introduction}
\renewcommand{\theequation}{1.\arabic{equation}} \setcounter{equation}{0}

Quantization of gravity is a key issue in modern theoretical
gravitational theory, since as a quantum field theory Einstein's general relativity with
Lorentz symmetry is unrenormalizable. A renormalizable candidate of
quantum gravity has been recently proposed by Ho\v{r}ava \cite{HL}, who
assumed that  the Lorentz symmetry is broken in the ultraviolet, so that
the anisotropic scalings between space and time are given by
 \bqn
 \label{INT1}
\textbf{x}\rightarrow \ell\textbf{x},~~~~~~~t\rightarrow \ell^z t\quad .
 \eqn
A power-counting renormalizable gravity theory must satisfy $z\geq
3$ in 3+1 dimensional spacetime. Such a theory is called the Ho\v{r}ava-Lifshitz gravity.

Ho\v{r}ava-Lifshitz theory has attracted the attention of many
theoretical physicists. However, it also faces several
problems, in particular arising from the spin-0 graviton. In order to solve such shortcomings,
a local U(1) gauge field $A$ is introduced\cite{US}. Meanwhile some recent works prove that the problems from
spin-0 graviton can be cancelled in Ho\v{r}ava-Lifshtiz gravity by means of the
U(1) gauge field $A$ \cite{Work}, and the post-newtonian
approximation is also satisfied \cite{PostNewtonian}.

On the other hand, recently, Hartnoll, Herzog and Horowitz
considered the AdS/CFT (Anti de Sitter/Conformal Field Theory)
correspondence principle to study the strongly correlated
condensed matter physics with the gravitational dual. They found
a correspondence between the instability of black string and the
second-order phase transition from normal to superconductor
state.

Subsequently, several authours generalized the idea to investigate holographic
superconductors of various black hole solutions \cite{everybody}. However, the
holographic superconductor models are under the framework of
spacetime with Lorentz symmetry. In this paper, we  
build a holographic superconductor model with non-relativistic
matter in static Ho\v{r}ava-Lifshtiz spacetime. The present research
could help understanding the stability properties in
Ho\v{r}ava-Lifshtiz spacetime.

We plan this paper as follows. In section II, we generalize the
model in general relativity to build a holographic superconductor
action in Ho\v{r}ava-Lifshitz spacetime, and focus on discussing the
correction from non-relativistic terms in section III. Then, in
section IV, the conductivity will be calculated, and section V
includes some conclusions and a summary.

\section{superconductor action in Ho\v{r}ava-Lifshitz gravity}
\renewcommand{\theequation}{2.\arabic{equation}} \setcounter{equation}{0}

In this section, we build a holographic superconductor model in
Ho\v{r}ava-Lifshitz gravity. We first analyze the
holographic superconductor in general relativity. In
\cite{Holographic}, Hartnoll, Herzog and Horowitz proposed the model
with the Lagrangian density
 \bqn
 \label{GRM1}
 {\cal L}_G=-\frac{1}{4}{\cal F}^{\mu\nu}{\cal
 F}_{\mu\nu}-\left|\nabla \bar{\Psi}-iq{\cal
 A}\bar{\Psi}\right|^2-V(|\bar{\Psi}|).
 \eqn
where $\bar{\Psi}$ is the scalar field, ${\cal A}_\nu$ is
electromagnetic four-potential in general relativity, and ${\cal
F}_{\mu\nu}=\partial_\mu {\cal A}_\nu-\partial_\nu {\cal A}_\mu$.
The holographic superconductor model also requests
 \bqn
 \label{GRM2}
V(|\bar{\Psi}|)=m^2\left|\bar{\Psi}\right|^2.
 \eqn
 Rewritting the scalar field $\bar{\Psi}$ as $\Psi e^{ip}$
\cite{Transform}, with real $\Psi$ and $p$, Eq.(\ref{GRM1})
becomes
 \bqn
 \label{GRM3}
 {\cal L}_G=-\frac{1}{4}{\cal F}^{\mu\nu}{\cal
 F}_{\mu\nu}-\nabla_\mu {\Psi}\nabla^\mu {\Psi}-V(|{\Psi}|)\nb\\
 -\left(q{\cal A}_\mu -\partial_\mu p\right)\left(q{\cal A}^\mu-\partial^\mu
p\right)\Psi^2.
 \eqn
We choose the gauge $p=0$ and get the action of the
simplest models in general relativity,
 \bqn
 \label{GRM4}
S_G=\int d^{4}x\sqrt{g^{(4)}}\left(\frac{1}{4}{\cal L}_G^E+2{\cal
L}_G^S-{\cal L}_G^C\right),
 \eqn
where $g^{(4)}$ is the determinant of the 4 dimensional metric
$g^{(4)}_{\mu\nu}$, while the electromagnetic, scalar and
coupling parts of the Lagrangian are, respectively, given by
 \bqn
 \label{GRM5}
{\cal L}_G^E&=&{\cal F}^{\mu\nu}{\cal F}_{\mu\nu},\nb\\
{\cal L}_G^S&=&-\frac{1}{2}\nabla_\mu {\Psi}\nabla^\mu {\Psi}-\frac{1}{2}V(|{\Psi}|),\nb\\
{\cal L}_G^C&=&q^2g^{(4)\mu\nu}{\cal A}_\mu{\cal A}_\nu\Psi^2\quad .
 \eqn
Next, we generalize the action in Ho\v{r}ava-Lifshitz theory, with
the Arnowitt-Deser-Misner metric
 \bqn
 \label{metric0}
ds^2=-N^2dt^2+g_{ij}\left(dx^i-N^idt\right)\left(dx^j-N^jdt\right).
 \eqn
The non-relativistic matter in Ho\v{r}ava-Lifshitz gravity was
proposed in \cite{Matter} and the Lagrangians of complex scalar
and electromagnetic fields are
 \bqn
 \label{HM1}
{\cal L}_H^E&=&\frac{2}{N^2}g^{ij}\left(F_{0i}-F_{ki}N^k\right)\left(F_{0j}-F_{lj}N^l\right)\nb\\
&&-F_{ij}F^{ij}-\beta_0-\beta_1a_iB^i-\beta_2B_iB^i-{\cal G}_E\quad ,\nb\\
{\cal
L}_H^S&=&\frac{1}{2N^2}\left|\partial_t{\Psi}-N^i\partial_i\Psi\right|^2-\frac{1}{2}\left|\partial\Psi\right|^2\nb\\
&&-\frac{1}{2}V(|{\Psi}|)+\alpha_2\left|\partial\Psi\right|^2-{\cal
H}_S\quad ,
 \eqn
where $F_{ij}=\partial_jA_i-\partial_iA_j$. The Ho\v{r}ava-Lifshitz
higher order corrections ${\cal G}_E$ and ${\cal H}_S$ are given by
 \bqn
 \label{HM2}
{\cal G}_E&=&\beta_3\left(B_iB^i\right)^2+\beta_4\left(B_iB^i\right)^3+\beta_5\left(\nabla_iB_j\right)\left(\nabla^iB^j\right)\nb\\
&&+\beta_6\left(B_iB^i\right)\left(\nabla_kB_j\right)\left(\nabla^kB^j\right)\nb\\
&&+\beta_7\left(\nabla_iB_j\right)\left(\nabla^iB^k\right)\left(\nabla^jB_j\right)\nb\\
&&+\beta_8\left(\nabla_i\nabla_jB_k\right)\left(\nabla^i\nabla^jB^k\right)\quad ,\nb\\
{\cal
H}_S&=&\alpha_3\left(\Psi\Delta\Psi\right)^2+\alpha_4\left(\Psi\Delta\Psi\right)^3+\alpha_5\Psi\Delta^2\Psi\nb\\
&&+\alpha_6\left(\Psi\Delta\Psi\right)\left(\Psi\Delta^2\Psi\right)+\alpha_7\Psi\Delta^3\Psi\quad ,
 \eqn
where $\alpha_i$ are arbitrary functions of $\Psi$ and $\beta_i$
arbitrary functions of $A_iA^i$. Nonetheless, we consider $\alpha_i$ and
$\beta_i$ as constants in this paper, what can be seen as a weak field approximation.
Moreover, $B^i=\frac{1}{2}\frac{\epsilon^{ijk}}{\sqrt{g}}F_{jk}$ with the
Levi-Civita symbol $\epsilon^{ijk}$. What we want to consider is
just the lower order terms of above equations: the higher order
terms ${\cal G}_E$ and ${\cal H}_S$ are ignored in this paper.

Now, let's construct the coupling between electromagnetic field and
scalar field. The simplest transformations are given by
 \bqn
 \label{TR1}
p_i&\rightarrow& p_i-qA_i\quad ,\nb\\
p_0&\rightarrow& p_0-qA_0\quad ,
 \eqn
where $A_i$ and $A_0$ satisfy the gauge invariant
 \bqn
 \label{TR2}
A_i&\rightarrow& A_i+\nabla_i\chi\quad ,\nb\\
A_0&\rightarrow& A_0-\partial_t{\chi}\quad .
 \eqn
Therefore, we make the replacement
 \bqn
 \label{TR3}
\nabla_i&\rightarrow&\nabla_i-iqA_i\quad ,\nb\\
\partial_0&\rightarrow& \partial_0-iqA_0\quad ,
 \eqn
and the complex scalar field is rewritten as
 \bqn
 \label{TR4}
\tilde{\cal
L}_H^S=\frac{1}{2N^2}\left|\partial_t{\Psi}-iqA_0\Psi-N^i\left(\partial_i\Psi-iqA_i\Psi\right)\right|^2\nb\\
-\left(\frac{1}{2}-\alpha_2\right)\left|\partial\Psi-iqA_i\Psi\right|^2-\frac{1}{2}V(|{\Psi}|)-\tilde{\cal
H}_S\quad ,\nb\\
 \eqn
where ${\cal H}_S$ is replaced by $\tilde{\cal H}_S$ with
$\partial_i\rightarrow\partial_i-iqA_i$.

Therefore, we builded a holographic superconductor model in
Ho\v{r}ava-Lifshitz gravity,
 \bqn
 \label{TR5}
S_H=\int dtd^3x N\sqrt{g}\left(\frac{1}{4}{\cal L}_H^E+2\tilde{\cal
L}_H^S\right)\quad .
 \eqn

Considering the relationship \cite{Matter}
 \bqn
 \label{TR6}
g^{(4)00}=-\frac{1}{N^2},~~~g^{(4)0i}=\frac{N^i}{N^2}\quad ,\nb\\
g^{(4)ij}=g^{ij}-\frac{N^iN^j}{N^2}\quad ,
 \eqn
eq.(\ref{TR5}) reduces into Eq.(\ref{GRM1}) when
$\alpha_i=\beta_i=0$.

\section{Numerical Results}
\renewcommand{\theequation}{3.\arabic{equation}} \setcounter{equation}{0}

In this paper, we aim at the holographic
superconductor in the low energy limit of Ho\v{r}ava-Lifshitz gravity,
while Schwarzschild spacetime is one of the solutions of the low
energy limit in Ho\v{r}ava-Lifshitz gravity\cite{Metric}. Thus, let us
consider the Schwarzschild spacetime
 \bqn
 \label{Metric1}
ds^2=-N^2(r)dt^2+\frac{dr^2}{f(r)}+r^2\left(dx^2+dy^2\right)\quad ,
 \eqn
with
 \bqn
 \label{Metric2}
N^2(r)=f(r)=r^2-\frac{r_0^3}{r}\quad ,
 \eqn
where we have chosen the AdS radius $L=1$. In the low energy case, we
consider the temperature of black hole as given by $T_h=\frac{3r_0}{4\pi}$. We
also set the simplest form for the mass parameter $m^2=-2+4\alpha_2$. Therefore, we derive
the field equations in this Ho\v{r}ava-Lifshitz spacetime,
 \bqn
 \label{FE1}
&&\Psi''+\left(\frac{2}{r}+\frac{f'}{f}\right)\Psi'+\left[\frac{2}{f}+\frac{\Phi^2}{(1-2\alpha_2)f^2}\right]\Psi=0,\nb\\
&&\Phi''+\frac{2}{r}\Phi'-\frac{2\Psi^2}{f}\Phi=0\quad .
 \eqn
At infinity, because $f(r)\rightarrow r^2$, we can get the boundary
condition for $\Psi$ and $\Phi$
 \bqn
 \label{BC1}
\Psi&=&\frac{\sqrt{2}\langle{\cal
O}_1\rangle}{r}+\frac{\sqrt{2}\langle{\cal
O}_2\rangle}{r^2}+\cdot\cdot\cdot,\nb\\
\Phi&=&\mu-\frac{\rho}{r}+\cdot\cdot\cdot.
 \eqn
Substituting the boundary condition into the main equations of the
holographic superconductor, we can use the shooting method to calculate
Eq.(\ref{FE1}) numerically. Note that the correction from
Ho\v{r}ava-Lifshitz gravity in Eq.(\ref{FE1}) is $\alpha_2$, so we
focus on studying the effect of $\alpha_2$.

We set $r_0=1$ in the definition of $f(r)$ and the charge of test particle as unit, $q=1$. The
critical temperature $T_c$ in ${\cal O}_1$ and ${\cal O}_2$ is
given in Table I.

\begin{table}[ht]
\caption{\label{TableI} Critical temperature $T_c$ with ${\cal O}_1$
and ${\cal O}_2$ respectively}
\begin{tabular}{c c c}
         \hline
~~~~~~~~~~$\alpha_2$~~~~~~~~~&~~~~~~~~$\langle{\cal
O}_1\rangle$~~~~~~~~~~~~&~~~~~~~~~$\langle{\cal O}_2\rangle$~~~~~~~~
        \\
        \hline
$0$&$0.2255\rho^{1/2}$&$0.1184\rho^{1/2}$
          \\
$0.1$&$0.2385\rho^{1/2}$&$0.1252\rho^{1/2}$
          \\
$0.2$&$0.2563\rho^{1/2}$&$0.1346\rho^{1/2}$
          \\
$0.3$&$0.2836\rho^{1/2}$&$0.1489\rho^{1/2}$
          \\
$0.4$&$0.3373\rho^{1/2}$&$0.1771\rho^{1/2}$
          \\
        \hline
\end{tabular}
\end{table}
The results show that the effect of $\alpha_2$ is to increase the critical
temperature, while the effect of $\langle{\cal O}_2\rangle$
is more obvious than the effect of $\langle{\cal O}_1\rangle$.

\begin{figure*}
\includegraphics[width=8cm]{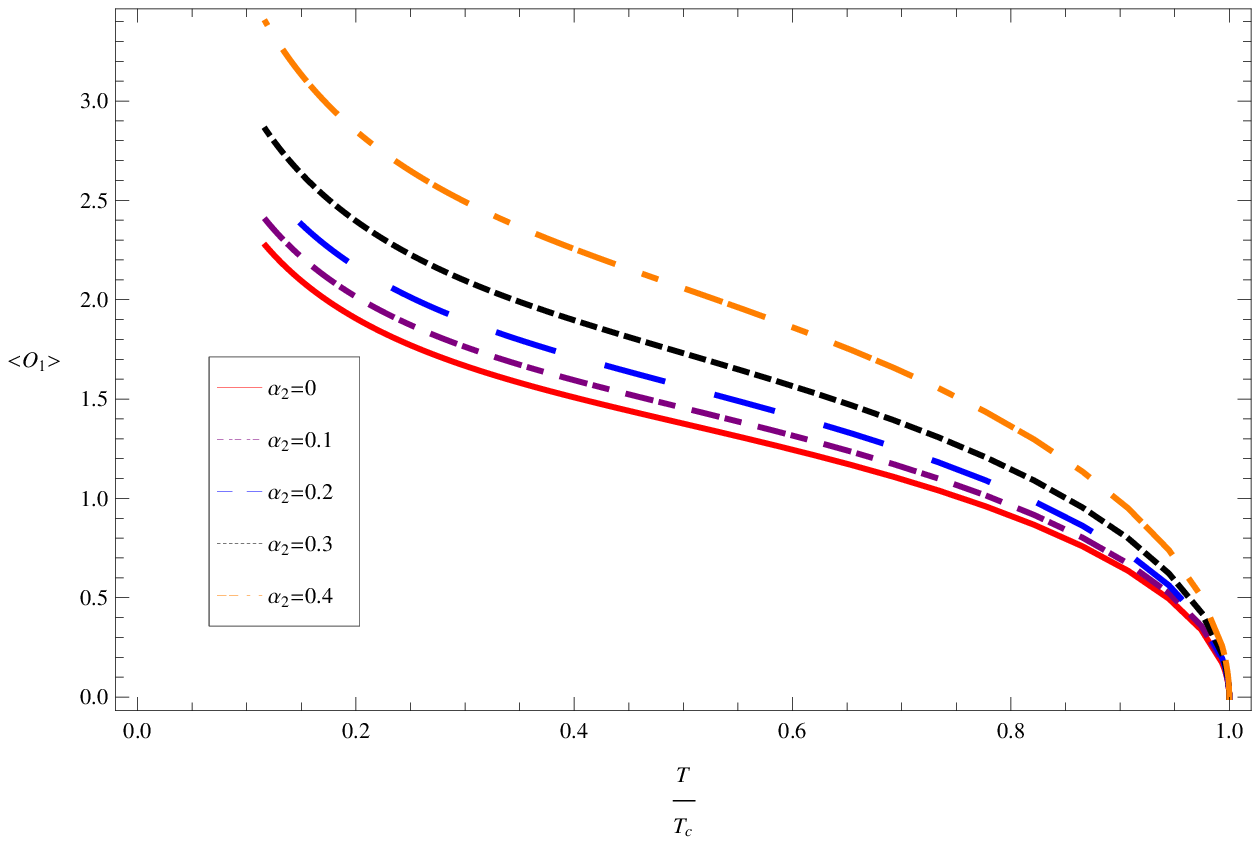}\includegraphics[width=8cm]{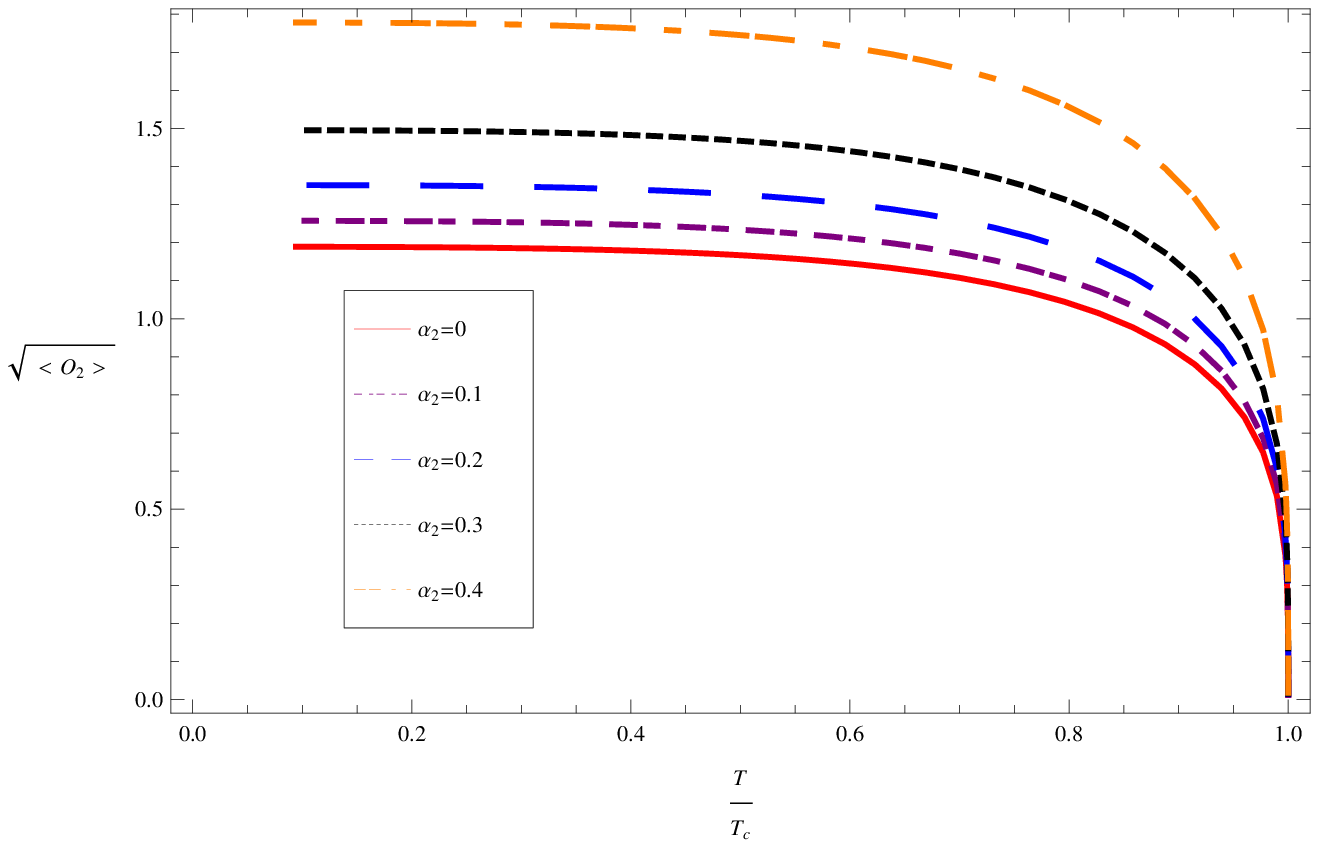}
\caption{The condensate as a function of the temperature for the
operators ${\cal O}_1$ and ${\cal O}_2$.} \label{figA}
\end{figure*}

Then, we draw the transition curve in FIG.\ref{figA}, and we find
that, as the $\alpha_2$ increases, the curved line gets higher.

\section{Conductivity}
\renewcommand{\theequation}{4.\arabic{equation}} \setcounter{equation}{0}
Here, we discuss the conductivity. Considering the
perturbed Maxwell field $A_i=\delta^x_ie^{-i\omega t}A_x(r)$ while
$A_0=0$, we obtain the equation
 \bqn
 \label{MW1}
&&A''_x(r)+\frac{8r^3\beta_2+f'(r)}{2r^4\beta_2+f(r)}A'_x(r)\nb\\
&&+\frac{\omega^2-2(1-2\alpha_2)f(r)\Psi^2}{\left(2r^4\beta_2+f(r)\right)f(r)}A_x(r)=0\quad ,\nb\\
 \eqn
but what we are interested in is the case $\beta_2=0$, in which case the above equation is
rewritten as
 \bqn
 \label{MW2}
A''_x+\frac{f'(r)}{f(r)}A'_x+\left[\frac{\omega^2}{f(r)^2}-\frac{2(1-2\alpha_2)\Psi^2}{f(r)}\right]A_x=0\quad .
 \eqn
We consider the low energy scale case, when
the boundary condition at $r=r_0$ requires
 \bqn
 \label{MW3}
A_x(r)\sim f(r)^{-\frac{\omega}{3r_0}}\quad ,
 \eqn
while the behavior of $A_x$ in the asymptotic AdS region is given by
 \bqn
 \label{MW4}
A_x(r)=A^{(0)}_x+\frac{A^{(1)}_x}{r}+\cdot\cdot\cdot\quad ,
 \eqn
and the definition of conductivity is \cite{Holographic}
 \bqn
 \label{MW5}
\sigma=-\frac{iA^{(1)}_x}{\omega A^{(0)}_x}\quad .
 \eqn

Using the above formulas, we draw the relation between
$\text{Re}(\sigma)$ and $\omega$. We find that the real part of
conductivity curved lines are lower as $\alpha_2$
increases. On the other hand, from the relations between
$-\text{Im}(\sigma)$ and $\omega$, we find that the imaginary part
of the conductivity  lines are higher as $\alpha_2$ increases.

\begin{figure*}
\includegraphics[width=8cm]{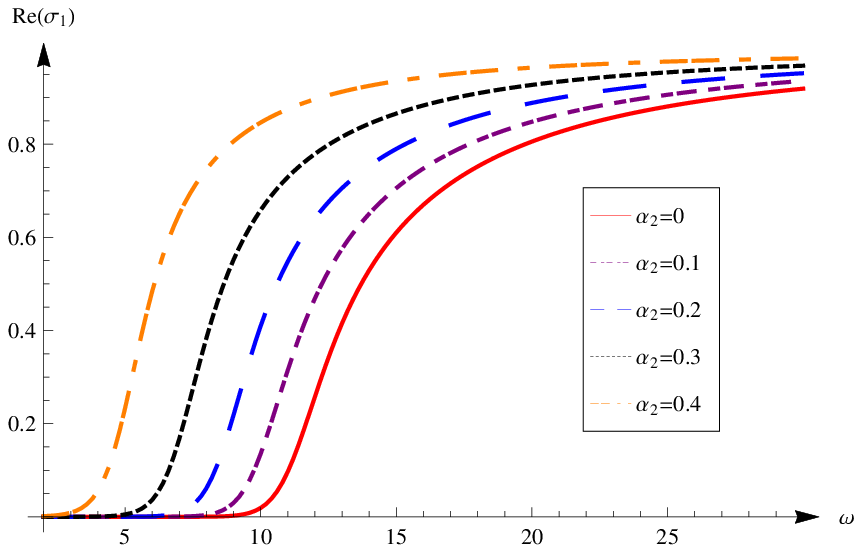}\includegraphics[width=8cm]{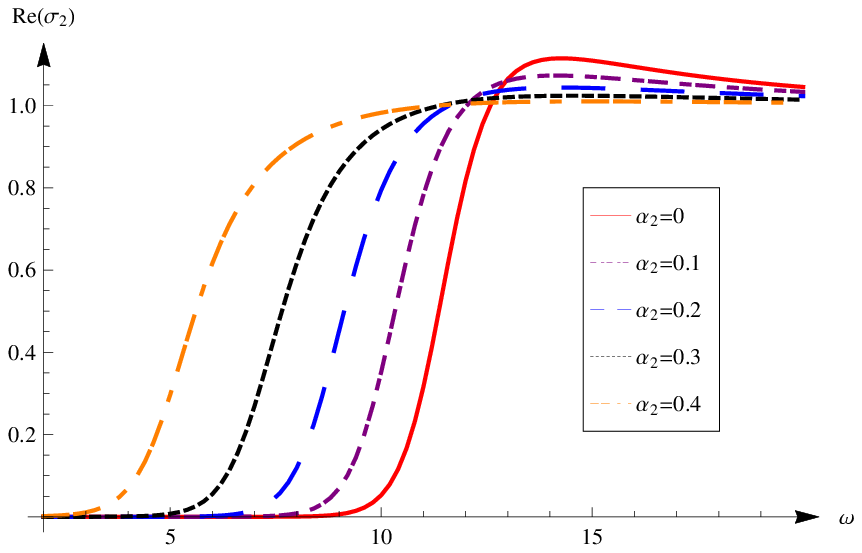}
\caption{The real part of conductivity for the two operators ${\cal
O}_1$ and ${\cal O}_2$.} \label{figB}
\end{figure*}

\begin{figure*}
\includegraphics[width=8cm]{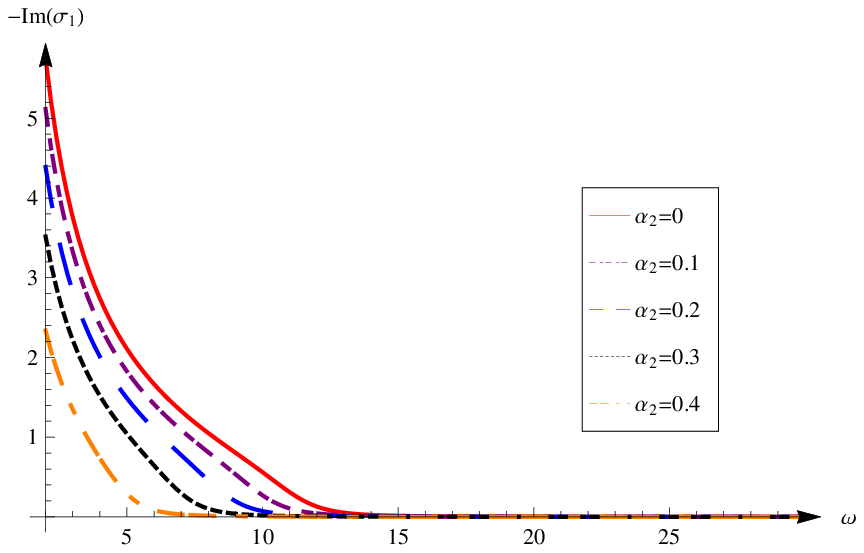}\includegraphics[width=8cm]{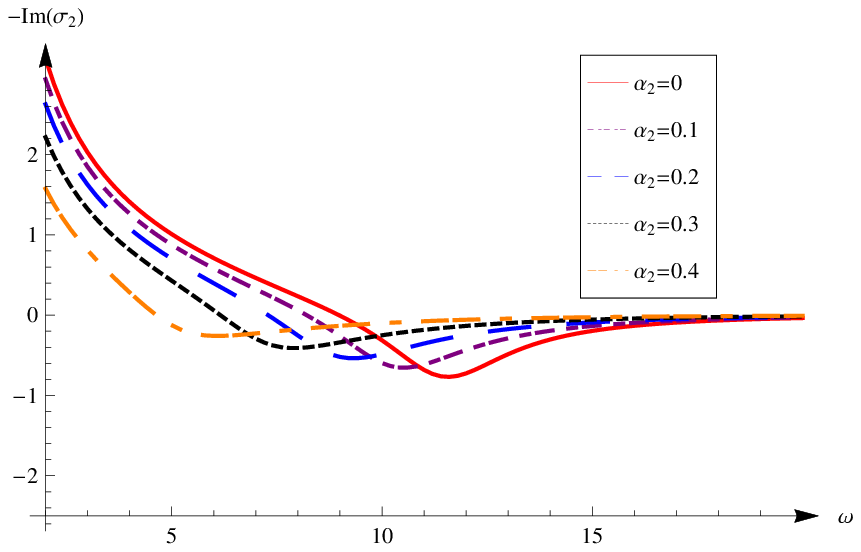}
\caption{ The imaginary part of conductivity for the two operators
${\cal O}_1$ and ${\cal O}_2$.} \label{figC}
\end{figure*}

\begin{figure*}
\includegraphics[width=8cm]{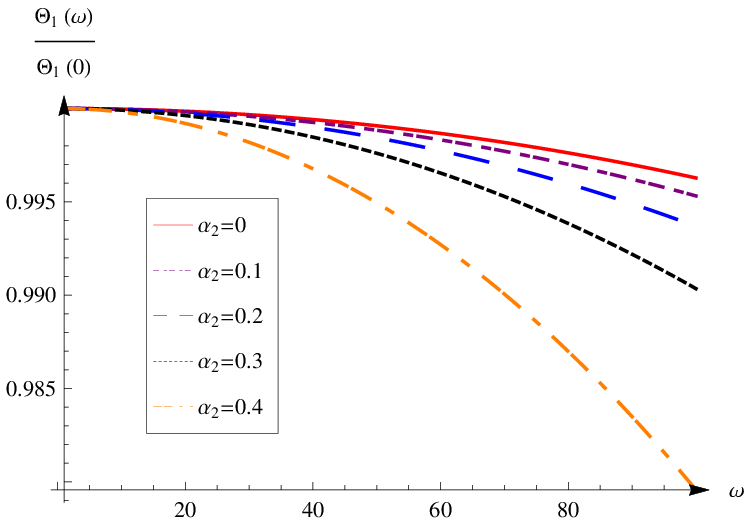}\includegraphics[width=8cm]{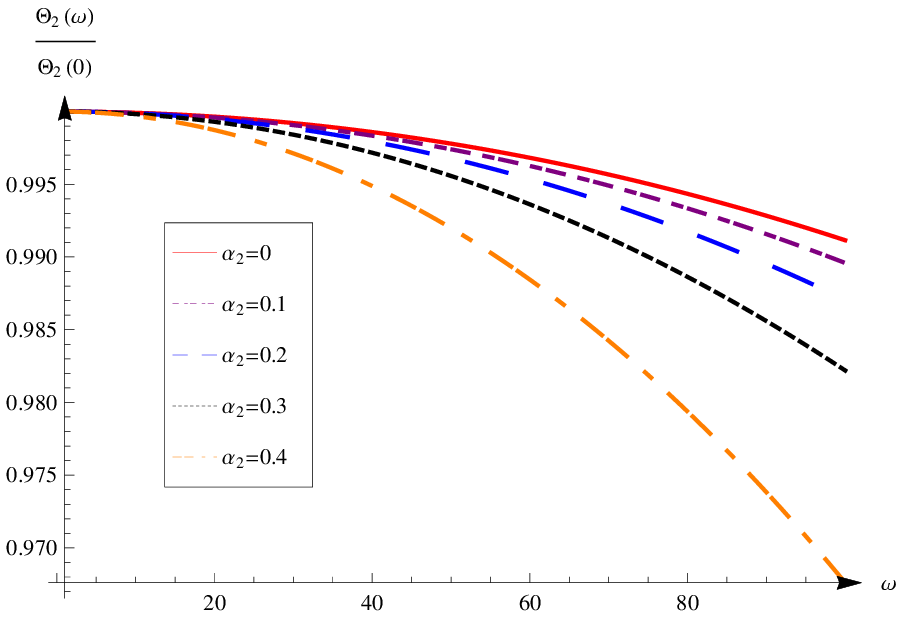}
\caption{$\frac{\Theta(\omega)}{\Theta(0)}$ for the two operators
${\cal O}_1$ and ${\cal O}_2$ in rotating spacetime, where
$\Theta(\omega)\equiv-\omega \text{Im}\sigma$} \label{figD}
\end{figure*}

Finally, we plot $\frac{\Theta(\omega)}{\Theta(0)}$ with small
$\omega$ (where $\Theta\equiv-\omega \text{Im} \sigma$), and FIG.
\ref{figD} shows that the $\frac{\Theta(\omega)}{\Theta(0)}$ goes to
a constant as $\omega$ is enough small, but the curves are lower as $\alpha_2$ increases. 
The values of $\Theta(0)$ are given in table II.

\begin{table}[ht]
\caption{\label{TableII} $\Theta_1(0)$ and $\Theta_2(0)$ with
$\alpha_2$ as $r_0\Psi(r_0)=5.8$}
\begin{tabular}{c c c}
         \hline
~~~~~~~~~~$\alpha_2$~~~~~~~~~&~~~~~~~~$\Theta_1(0)$~~~~~~~~~~~~&~~~~~~~~~$\Theta_2(0)$~~~~~~~~
        \\
        \hline
$0$&$11.6798$&$6.6225$
          \\
$0.1$&$10.4294$&$6.1340$
          \\
$0.2$&$9.0088$&$5.5426$
          \\
$0.3$&$7.3211$&$4.7786$
          \\
$0.4$&$5.1131$&$3.6468$
          \\
        \hline
\end{tabular}
\end{table}

\section{Conclusion}
\renewcommand{\theequation}{5.\arabic{equation}} \setcounter{equation}{0}

We built the simplest holographic superconductor model in the
low energy limit of Ho\v{r}ava-Lifshitz gravity and studied
the property of transition near the critical temperature $T_c$ in
static Ho\v{r}ava-Lifshitz spacetime. We found that the correction comes
from the $\alpha_2\Psi\Delta\Psi$ term in 3+1 dimensional static
spacetime. The study also shows that the correction of
$\langle{\cal O}_2\rangle$ is more obvious than the effect on
$\langle{\cal O}_1\rangle$. From Eq.(\ref{FE1}), if we make the
transformation
$\frac{\Phi}{\sqrt{1-2\alpha_2}}\rightarrow\tilde\Phi$, we find the
equations of superconductors will give the same results for any
constant $\alpha_2$, but $\alpha_2$ can modify the conductivity in
Section IV.

What we considered is just the simplest case, but it is possible
that the correction comes from ${\cal H}_S$ and ${\cal G}_E$ as the
black hole is charged or rotated. On the other hand, our work proves
that it should introduce a $U(1)$ symmetrical field to avoid the
difficulties from spin-0 graviton in Ho\v{r}ava-Lifshitz theory
\cite{Work}\cite{PostNewtonian}, so it is more meaningful to
research the holographic superconductor in general case with a $U(1)$
symmetric field. 

The effect of the constant  $\alpha_2$ is to decrease the value of conductivity and lower the
superconductor effect (see table II). However, we did not see and insulator effect, a case
in which we have to consider large values of  $\alpha_2$, which then might be field dependent.

\section*{\bf Acknowledgements}

This work is supported in part  by FAPESP No. 2012/08934-0 (EA, KL);  CNPq  (EA, KL);   DOE  Grant, DE-FG02-10ER41692 (AW);
Ci\^encia Sem Fronteiras, No. 004/2013 - DRI/CAPES (AW); and NSFC No. 11375153 (AW).

\onecolumngrid

\end{document}